\documentclass[aps,prc,preprint,amsmath,showpacs,superscriptaddress,nofootinbib]{revtex4}
\usepackage{graphicx,epsfig,epstopdf}
\newcommand{\beqy}{\begin{eqnarray}}
\newcommand{\eeqy}{\end{eqnarray}}
\newcommand{\bmlet}{\begin{subequations}}
\newcommand{\emlet}{\end{subequations}}

\begin{document}

\textwidth 16.2 cm
\oddsidemargin -.54 cm
\evensidemargin -.54 cm

\def\gsimeq{\,\,\raise0.14em\hbox{$>$}\kern-0.76em\lower0.28em\hbox  
{$\sim$}\,\,}  
\def\lsimeq{\,\,\raise0.14em\hbox{$<$}\kern-0.76em\lower0.28em\hbox  
{$\sim$}\,\,}

\title{Spin and spin-isospin instabilities in asymmetric nuclear matter at zero and finite 
temperatures using Skyrme functionals}
\author{N.~Chamel}
\affiliation{Institut d'Astronomie et d'Astrophysique, CP-226, Universit\'e
Libre de Bruxelles, 1050 Brussels, Belgium}
\author{S.~Goriely}
\affiliation{Institut d'Astronomie et d'Astrophysique, CP-226,
Universit\'e Libre de Bruxelles, 1050 Brussels, Belgium}

\date{\today}

\begin{abstract}
Self-consistent mean field methods based on 
phenomenological Skyrme effective interactions are known to exhibit 
spurious spin and spin-isospin instabilities both at zero and 
finite temperatures when applied to homogeneous nuclear matter
at the densities encountered in neutron stars and in supernova cores. 
The origin of these instabilities is revisited in the framework of the
nuclear energy density functional theory and a simple prescription is proposed to 
remove them. The stability of several Skyrme parametrizations is 
reexamined. 
\end{abstract}


\maketitle

\section{Introduction}

The self-consistent mean-field method with Skyrme effective interactions has 
been very successful in describing the structure and the dynamics of 
medium-mass and heavy nuclei~\cite{bhr03}. These interactions have been also 
widely applied to the description of extreme astrophysical environments 
such as neutron stars and supernova cores. Actually very soon after Skyrme~\cite{sky59} 
introduced his eponymous effective interaction, Cameron~\cite{cam59} applied 
it to calculate the structure of neutron stars. Assuming that neutron stars were 
made only of neutrons, he found that their maximum mass was significantly higher 
than the Chandrasekhar mass limit. His work thus brought support to the scenario 
of neutron star formation from the catastrophic gravitational collapse of massive 
stars in supernova explosions, as proposed much earlier by Baade and Zwicky~\cite{bz33}. 
The interior of neutron stars is highly neutron rich but contains also a non-negligible 
amount of protons, leptons and possibly other particles. However microscopic calculations 
in uniform infinite nuclear matter using bare nucleon-nucleon potentials have been usually 
restricted to symmetric nuclear matter (SNM) and pure neutron matter (NeuM). Even though 
effective interactions are phenomenological, they can provide a convenient interpolation of 
realistic calculations to determine the equation of state of neutron star cores. Mean-field 
calculations can be easily extended to finite temperatures and can thus be also used to 
describe the hot nuclear matter found in supernova cores and protoneutron stars. 
Moreover, the mean-field method allows a consistent and tractable treatment of both 
homogeneous matter and inhomogeneous matter (e.g. neutron star crusts~\cite{lrr}) with a reduced 
computational cost. This opens the way to a unified description of all regions of neutron 
stars and supernova cores~\cite{sto07}. 

Nevertheless the application of these effective forces to nuclear matter at high densities 
has been limited by the occurrence of spurious instabilities~\cite{mar02, agra04}. In particular, 
Skyrme forces predict a spontaneous transition to a spin-polarized phase when the density exceeds 
a critical threshold which depends on the isospin asymmetry~\cite{vida84,isa06,gar08,mar09a,cgp09}. 
Besides, it is found that for some forces the energy density of the spin-polarized phase decreases 
with increasing density. In this case, the phase transition is accompanied by a catastrophic collapse~\cite{kw94}, 
which is contradicted by the existence of neutron stars (note however that observations alone do not exclude the 
possibility of a ferromagnetic core inside neutron stars, see for instance Refs.~\cite{hae96, kut99}). 
Moreover, the critical density predicted within the Skyrme formalism generally decreases with temperature due 
to an anomalous behavior of 
the entropy, which is larger in the spin-ordered phase than in the unpolarized phase~\cite{rios05, isa10}. 
This instability can strongly affect the neutrino propagation in hot dense nuclear matter~\cite{red99, na99, mar09a} 
which is believed to play an important role in the supernova explosion mechanism and in the evolution of 
protoneutron stars~\cite{pra97}. However, no such spin-polarized phase transition is found by microscopic 
calculations using realistic nucleon-nucleon potentials. Indeed several calculations based on different 
methods, such as the lowest-order constrained variational method~\cite{bord08a,bord08b,big09,mod10,big10}, the 
Brueckner-Hartree-Fock method~\cite{vid02, zuo02, bomb06}, the auxiliary field diffusion Monte Carlo 
method~\cite{fan01} and the Dirac-Brueckner-Hartree-Fock method~\cite{sam07}, show that nuclear matter 
remains unpolarized well above the nuclear saturation density $\rho_0$ both at zero and finite temperatures. 

The prediction of spin-ordering in nuclear matter is one of the main deficiencies of the mean-field 
method with effective forces. Different extensions of the standard Skyrme force have been 
recently proposed in order to prevent these phase transitions at zero temperature~\cite{mar09a, cgp09}. 
In this paper, the origin of the spin and spin-isospin instabilities is revisited in the more general framework of 
the nuclear energy density functional (EDF) theory (see for instance Ref.~\cite{drut10} for a review) and 
a simpler prescription is proposed to ensure stability of dense nuclear matter for any degree
of spin and spin-isospin polarizations and for any temperature. The paper is organized as follows. 
The Skyrme functionals that we consider here are defined in Section~\ref{sec2}. 
Section~\ref{sec3} is devoted to the discussion about the stability of 
nuclear matter. Several Skyrme functionals are reexamined in Section~\ref{sec4}.

\section{Skyrme functionals}
\label{sec2}

The nuclear EDFs that we consider here are of the form
\beqy
\label{1}
E=E_{\rm kin}+E_{\rm Coul}+E_{\rm Sky}\quad ,
\eeqy
where $E_{\rm kin}$ is the kinetic energy, $E_{\rm Coul}$ is the Coulomb energy 
and $E_{\rm Sky}$ is a functional of the 
local densities and currents ($q=n,p$ for neutron, proton respectively): the density 
$\rho_q$, the current density $\pmb{j_q}$, the kinetic density $\tau_q$, the spin density $\pmb{s_q}$, 
the spin kinetic density $\pmb{T_q}$ and the spin-current tensor $J_{q,\mu\nu}$
(see for instance Ref.~\cite{bhr03} for precise definitions). 
It is convenient to introduce the isospin index $t=0,1$ for isoscalar and isovector quantities respectively.
Isoscalar quantities (also written without any subscript) are sums over neutrons and protons (e.g. 
$\rho_0=\rho=\rho_n+\rho_p$) while isovector quantities are differences between neutrons and protons 
(e.g. $\rho_1=\rho_n-\rho_p$). The Skyrme functional $E_{\rm Sky}$ is then given by
\bmlet 
\beqy
\label{2a}
E_{\rm Sky}=\int{\rm d}^3\pmb{r}\,\mathcal{E}_{\rm Sky}(\pmb{r}),\quad \mathcal{E}_{\rm Sky}=\sum_{t=0,1}(\mathcal{E}_t^{\rm even}+\mathcal{E}_t^{\rm odd})\quad, 
\eeqy
\beqy
\label{2b}
\mathcal{E}_t^{\rm even}=C_t^\rho \rho_t^2+C_t^{\Delta\rho}\rho_t\Delta\rho_t+C_t^\tau\rho_t\tau_t
+C_t^{\nabla J}\rho_t\nabla\cdot\pmb{J_t}+C_t^J \sum_{\mu,\nu}J_{t,\mu\nu}J_{t,\mu\nu}\quad ,
\eeqy
\beqy
\label{2c}
\mathcal{E}_t^{\rm odd}=C_t^s s_t^2+C_t^{\Delta s}\pmb{s_t}\cdot \Delta \pmb{s_t}+C_t^T \pmb{s_t}\cdot\pmb{T_t}
+C_t^j j_t^2+C_t^{\nabla j} \pmb{s_t}\cdot\nabla\times\pmb{j_t}\quad .
\eeqy
\emlet 
The spin current vector is defined by $J_{t\kappa}=\sum_{\mu,\nu}\varepsilon_{\kappa\mu\nu}J_{t,\mu\nu}$, where 
$\varepsilon_{\kappa\mu\nu}$ is the Levi-Civita tensor. 
The so-called ``time-even'' part $\mathcal{E}_t^{\rm even}$ (``time-odd'' part $\mathcal{E}_t^{\rm odd}$) contains 
only even (odd) densities and currents with respect to time reversal. 

The coupling ``constants'' $C_t^\rho$ and $C_t^s$ generally depend on the isoscalar density $\rho=\rho_n+\rho_p$ as follows
\bmlet 
\beqy
\label{3a}
C_t^\rho=a_t^\rho + b_t^\rho \rho^\alpha\quad ,
\eeqy
\beqy
\label{3b}
C_t^s=a_t^s + b_t^s \rho^\alpha\quad .
\eeqy
\emlet 
Moreover, local gauge invariance~\cite{eng75,doba95} imposes the following relations
\beqy
\label{4}
C_t^j=-C_t^\tau\, , \quad C_t^J=-C_t^T\, ,\quad C_t^{\nabla j}=C_t^{\nabla J}\quad .
\eeqy
Historically the type of functionals given by Eqs.(\ref{2a})--(\ref{2c})
were obtained from the Hartree-Fock approximation using effective zero-range interactions 
of the Skyrme type~\cite{bhr03,sto07}
\beqy
\label{5}
v_{i,j} & = & 
t_0(1+x_0 P_\sigma)\delta({\pmb{r}_{ij}})
+\frac{1}{2} t_1(1+x_1 P_\sigma)\frac{1}{\hbar^2}\left[p_{ij}^2\,
\delta({\pmb{r}_{ij}}) +\delta({\pmb{r}_{ij}})\, p_{ij}^2 \right]\nonumber\\
& &+t_2(1+x_2 P_\sigma)\frac{1}{\hbar^2}\pmb{p}_{ij}\cdot\delta(\pmb{r}_{ij})\,
 \pmb{p}_{ij}
+\frac{1}{6}t_3(1+x_3 P_\sigma)\rho(\pmb{r})^\alpha\,\delta(\pmb{r}_{ij})
\nonumber\\
& &+\frac{\rm i}{\hbar^2}W_0(\pmb{\hat\sigma_i}+\pmb{\hat\sigma_j})\cdot
\pmb{p}_{ij}\times\delta(\pmb{r}_{ij})\,\pmb{p}_{ij}  \quad ,
\eeqy
where $\pmb{r}_{ij} = \pmb{r}_i - \pmb{r}_j$, $\pmb{r} = (\pmb{r}_i + 
\pmb{r}_j)/2$, $\pmb{p}_{ij} = - {\rm i}\hbar(\pmb{\nabla}_i-\pmb{\nabla}_j)/2$
is the relative momentum, $P_\sigma$ is the two-body spin-exchange 
operator. The relations between the coupling constants in Eqs.~(\ref{2b}) and (\ref{2c})
and the parameters of the effective force in Eq.~(\ref{5}), can be found 
for instance in Appendix A of Ref.~\cite{bhr03}. 

Kutschera and W\'ojcik~\cite{kw94} pointed out that for some Skyrme forces not only is the ground state of NeuM 
polarized, but also the energy density of polarized NeuM decreases with increasing density. 
However such a catastrophic ferromagnetic collapse is ruled out by neutron star observations. The origin of this 
singular behavior can be traced back to the parameters $t_2$ and $x_2$ of the Skyrme force. In particular, 
the authors of Ref.~\cite{kw94} found that in order to prevent a ferromagnetic collapse of NeuM, the 
parameters of Skyrme forces must satisfy the following inequality  
\beqy
\label{6}
t_2(1+x_2)\geq 0\quad .
\eeqy
This constraint was taken into account to construct the Saclay-Lyon Skyrme parametrizations~\cite{sly4}, 
which were fitted with the parameter $x_2=-1$. 
These forces which were specifically developed for 
astrophysics, have been widely used in neutron star studies. However it has been found that these forces
predict various transitions to spin-ordered phases in nuclear matter~\cite{rios05,isa06,mar09a,cgp09,isa10} 
even though Eq.~(\ref{6}) was enforced. Actually this is a general feature of standard Skyrme forces~\cite{mar02, agra04}. 
We will now reexamin this issue in the framework of the nuclear EDF theory.

\section{Stability of nuclear matter}
\label{sec3}

Let us consider the case of static uniform (possibly polarized) infinite isospin asymmetric nuclear matter. 
The Skyrme energy density, Eq.~(\ref{2a}), thus reduces to 
\beqy
\label{7}
\mathcal{E}_{\rm Sky}&=&\sum_{t=0,1}(C_t^\rho \rho_t^2+C_t^\tau\rho_t\tau_t
+C_t^s s_t^2+C_t^T \pmb{s_t}\cdot\pmb{T_t})\quad .
\eeqy
Let us choose the spin-quantization axis so that the only non-vanishing components 
of the spin density $\pmb{s_q}$ and the spin kinetic density $\pmb{T_q}$ are
along the z-axis. For brevity we will simply write $s_q$ and $T_q$ instead of $s_{qz}$ and 
$T_{qz}$. In the following we will neglect the anisotropies induced by the polarization of 
matter~\cite{dab76}. 
Introducing the density $\rho_{q\sigma}$ of nucleons with spins $\sigma=\uparrow,\downarrow$
and the kinetic density of polarized nucleons defined by 
\beqy
\label{8}
\tau_{q\sigma}=\frac{3}{5}(6\pi^2)^{2/3}\rho_{q\sigma}^{5/3}\quad ,
\eeqy
the spin density and the spin kinetic density can now be expressed as
\beqy
\label{9}
s_q=\rho_{q\uparrow}-\rho_{q\downarrow}\quad , 
\eeqy
\beqy
\label{10}
T_q=\tau_{q\uparrow}-\tau_{q\downarrow}\quad .
\eeqy
In fully polarized NeuM with all spins up ($\rho=\rho_{n\uparrow}$), Eq.~(\ref{7}) reduces to
\beqy
\label{11}
\mathcal{E}^{\rm pol}_{\rm NeuM}=\left[\frac{\hbar^2}{2M_n}+(C_0^\tau+C_1^\tau+C_0^T+C_1^T)\rho\right]\tau_{n\uparrow}
+(C_0^\rho+C_1^\rho+C_0^s+C_1^s)\rho^2\, .
\eeqy
If the energy density is calculated from a Skyrme force in the Hartree-Fock approximation, 
we find 
\beqy\label{12}
C_0^\rho+C_1^\rho+C_0^s+C_1^s=0\quad ,
\eeqy
\beqy\label{13}
C_0^\tau+C_1^\tau+C_0^T+C_1^T=\frac{1}{2}t_2(1+x_2) \quad ,
\eeqy
so that Eq.~(\ref{11}) reduces to
\beqy
\label{14}
\mathcal{E}^{\rm pol}_{\rm NeuM}=\left[\frac{\hbar^2}{2M_n}+\frac{1}{2}t_2(1+x_2)\rho\right]\tau_{n\uparrow}\quad .
\eeqy
Eq.~(\ref{12}) is a consequence of the Pauli exclusion principle and the zero 
range of the Skyrme interaction. Actually as will be shown elsewhere, Eq.~(\ref{12}) 
must still be satisfied for nuclear EDFs that are not obtained from effective forces 
in order to prevent self-interactions. 
The constraint of Kutschera and W\'ojcik~\cite{kw94}, Eq.~(\ref{6}), can thus 
be more generally written as
\beqy
\label{15}
C_0^\tau+C_1^\tau+C_0^T+C_1^T\geq 0\quad .
\eeqy
If this inequality is violated, $\mathcal{E}^{\rm pol}_{\rm NeuM}$ decreases
with increasing density thus leading to a ferromagnetic collapse. 

It is instructive to rewrite Eq.~(\ref{11}) as 
\beqy
\label{16}
\mathcal{E}^{\rm pol}_{\rm NeuM}=\frac{\hbar^2}{2M_{n\uparrow}^*}\tau_{n\uparrow}\quad ,
\eeqy
where we have introduced the effective mass of a nucleon in a spin state $\sigma$ defined by 
\beqy\label{17}
\frac{\hbar^2}{2M_{q\sigma}^*}=\frac{\partial\mathcal{E}}{\partial\tau_{q\sigma}}=\frac{\hbar^2}{2M_q^*}\pm\left[s(C_0^T-C_1^T)+2 s_q C_1^T\right]\quad ,
\eeqy
with $+$($-$) for spin up (spin down respectively), and $M_q^*$ is the usual effective mass 
given by 
\beqy\label{18}
\frac{\hbar^2}{2M_q^*}=\frac{\partial\mathcal{E}}{\partial\tau_{q}} = \frac{\hbar^2}{2M_q}+\left[(C_0^\tau-C_1^\tau)\rho+2\rho_q C_1^\tau\right]\quad .
\eeqy
It can be easily seen that in fully polarized NeuM, the effective mass reduces to
\beqy\label{19}
\frac{\hbar^2}{2M_{n\uparrow}^*}&=&\frac{\hbar^2}{2M_n}+(C_0^\tau+C_1^\tau+C_0^T+C_1^T)\rho \nonumber\\
&=& \frac{\hbar^2}{2M_n^*}+(C_0^T+C_1^T)\rho \nonumber\\
&=& \frac{\hbar^2}{2M_n}+t_2(1+x_2)\rho\quad 
\eeqy
so that Eq.~(\ref{16}) coincides with Eq.~(\ref{14}). Setting $x_2=-1$ as in the Saclay-Lyon Skyrme
forces~\cite{sly4} therefore implies that the effective mass of polarized neutrons is equal to the bare mass. 

We have seen that the constraint of Ref.~\cite{kw94} is equivalent to the requirement that the effective 
mass of polarized neutrons remains always positive. However the ground state of NeuM (and more generally 
that of isospin asymmetric nuclear matter) could still be polarized as shown below.

\subsection{Landau stability criterion}
\label{sec3A}
The stability of unpolarized homogeneous nuclear matter with respect to 
spin and spin-isospin polarizations has been generally addressed using 
the Landau Fermi-liquid theory (see e.g. Ref.~\cite{FL}). In this 
theory, the elementary excitations of the liquid at low temperatures are 
described in terms of quasiparticles which are in one-to-one correspondence 
with single-particle states of the non-interacting Fermi gas. 
Any \emph{small} change 
$\delta\tilde{n}(\pmb{k})$ in the distribution function of quasiparticles 
with wave vector $\pmb{k}$ leads to a change $\delta\mathcal{E}$ in the 
energy density, which can be expressed (up to second order) as 
\beqy
\label{20}
\delta\mathcal{E}=\int \frac{{\rm d}^3k}{(2\pi)^3}\, \varepsilon({\pmb{k}}) 
\delta\tilde{n}(\pmb{k})
+\frac{1}{2}\int\frac{{\rm d}^3k}{(2\pi)^3}\int\frac{{\rm d}^3k^\prime}{(2\pi)^3}\, 
v(\pmb{k},\pmb{k^\prime})\delta\tilde{n}(\pmb{k})\delta\tilde{n}(\pmb{k^\prime})
\eeqy
where $\varepsilon({\pmb{k}})$ is the energy of a quasiparticle with 
wave vector $\pmb{k}$ and $v(\pmb{k},\pmb{k^\prime})$ is the residual interaction 
between quasiparticles with wave vectors $\pmb{k}$ and $\pmb{k^\prime}$. 

In pure NeuM, the residual interaction (neglecting tensor interaction) can be expressed as
\beqy
\label{21}
v^{\rm NeuM}(\pmb{k},\pmb{k^\prime})=\frac{1}{N}\left[F^{\rm NeuM}(\pmb{k},\pmb{k^\prime})
+G^{\rm NeuM}(\pmb{k},\pmb{k^\prime}) \pmb{\hat\sigma}\cdot\pmb{\hat\sigma^\prime}\right]
\eeqy
where $N$ is the density of states at the Fermi surface given by 
\beqy
\label{22}
N=\frac{M^*_n k_{\rm F}}{\hbar^2\pi^2}\quad ,
\eeqy
with $k_{\rm F}=(3\pi^2\rho)^{1/3}$. We have also introduced the Pauli 
matrices $\pmb{\hat\sigma}$ and $\pmb{\hat\sigma^\prime}$ in order to take into account the spin 
of the quasiparticles. Small perturbations involve only quasiparticles at the Fermi surface, 
i.e. with $k=k^\prime=k_{\rm F}$. We can thus expand each term in the residual 
interaction in Legendre polynomials $P_\ell(\cos\theta)$ where $\theta$ 
is the angle between $\pmb{k}$ and $\pmb{k^\prime}$. For instance, 
\beqy
\label{23}
F^{\rm NeuM}(\pmb{k},\pmb{k^\prime})=\sum_{\ell=0}^{+\infty} F^{\rm NeuM}_\ell P_\ell(\cos\theta)
\eeqy
where $F^{\rm NeuM}_\ell$ are dimensionless Landau parameters. Similarly, we can define
Landau parameters $G^{\rm NeuM}_\ell$. For the Skyrme functional, the only 
non-zero Landau parameters are of order $\ell=0$ and $\ell=1$. 
The stability of the initial state is ensured if any change in 
the energy per particle $e\equiv\mathcal{E}/\rho$ is positive. This condition 
leads to Landau's criterion
\bmlet 
\beqy\label{24a}
F_\ell^{\rm NeuM} > -(2\ell+1) \quad ,
\eeqy
\beqy\label{24b}
G_\ell^{\rm NeuM} > -(2\ell+1)  \quad .
\eeqy
\emlet
In particular, the condition on $G_0^{\rm NeuM}$ guarantees that NeuM
is stable against small fluctuations of the (isoscalar) spin polarization 
$I_\sigma=s_0/\rho=(\rho_\uparrow-\rho_\downarrow)$. This can be seen 
by expanding the energy per particle up to second order in $I_\sigma$
\beqy
\label{25}
e(I_\sigma)\simeq e(0)+\frac{1}{2}
\frac{\partial^2 e}{\partial I_\sigma^2}\biggr\vert_{I_\sigma = 0} I_\sigma^2
\eeqy
with
\beqy
\label{26}
\frac{\partial^2 e}{\partial I_\sigma^2}\biggr\vert_{I_\sigma = 0} = 
\frac{\hbar^2 k_{\rm F}^2}{3 M_n^*}(1+G_0^{\rm NeuM})\quad .
\eeqy
The first order term vanishes because of the requirement that the unpolarized phase 
be an equilibrium state. 

Using the Skyrme functional, we find
\beqy
\label{27}
G_0^{\rm NeuM}=2N\Biggl[C_0^s+C_1^s+k_{\rm F}^2(C_0^T+C_1^T)\Biggr]\quad .
\eeqy
Now if the Skyrme functional is fitted to a realistic equation of state of NeuM~\cite{cgp09}, 
we find that $C_0^\rho+C_1^\rho\leq 0$, which according to Eq.~(\ref{12}) implies that 
\beqy
\label{29}
C_0^s+C_1^s\geq 0\quad . 
\eeqy
Ferromagnetic instabilities are therefore mainly due to the coupling constants $C_t^T$. 
In order to fulfill the Landau's stability condition $G_0^{\rm NeuM}>-1$ at 
any density, we must have\footnote{This inequality is not strictly
required if the coefficients $C_t^s$ are allowed to depend on the density according to Eqs.~(\ref{3a})
and (\ref{3b}) and the term in $(b_0^s+b_1^s)\rho^\alpha$ dominates at high density. However for modern 
Skyrme parametrizations such a situation does not arise because $3\alpha<2$.}
\beqy
\label{28}
C_0^T+C_1^T\geq 0\quad .
\eeqy

The absence of a ferromagnetic transition in NeuM does not generally forbid the occurence of 
spin-ordered phases in asymmetric nuclear matter. Let us consider in particular SNM. 
The most general form of the residual interaction (neglecting tensor interaction) can be expressed as
\beqy
\label{30}
v^{\rm SNM}(\pmb{k},\pmb{k^\prime})=\frac{1}{N_0}\left[F(\pmb{k},\pmb{k^\prime})+F^\prime(\pmb{k},\pmb{k^\prime})
 \pmb{\hat\tau}\cdot\pmb{\hat\tau^\prime}+G(\pmb{k},\pmb{k^\prime}) \pmb{\hat\sigma}\cdot\pmb{\hat\sigma^\prime}
+G^\prime(\pmb{k},\pmb{k^\prime}) \pmb{\hat\sigma}\cdot\pmb{\hat\sigma^\prime}\pmb{\hat\tau}\cdot\pmb{\hat\tau^\prime}\right]
\eeqy
where $N_0$ is the density of states at the Fermi surface given by
\beqy
\label{31}
N_0=\frac{2 M^*_s k_{{\rm F}0}}{\hbar^2\pi^2}\quad ,
\eeqy
with $k_{{\rm F}0}=(3\pi^2\rho/2)^{1/3}$ and $M_s^*$ is the isoscalar effective mass
defined by 
\beqy
\label{32}
\frac{M}{M_s^*}=1+\frac{2M}{\hbar^2}C_0^\tau\rho\, , \hskip 0.5cm \frac{2}{M}=\frac{1}{M_n}+\frac{1}{M_p}\quad .
\eeqy
We have also introduced the Pauli matrices $\pmb{\hat\tau}$, $\pmb{\hat\tau^\prime}$ in order to take into 
account the isospin of the quasiparticles. As before, we can define dimensionless Landau parameters
$F_\ell$, $F^\prime_\ell$, $G_\ell$ and $G_\ell^\prime$. The Landau's stability
conditions are in this case given by
\bmlet 
\beqy\label{33a}
F_\ell > -(2\ell+1) \quad ,
\eeqy
\beqy\label{33b}
F^\prime_\ell > -(2\ell+1)  \quad ,
\eeqy
\beqy\label{33c}
G_\ell > -(2\ell+1)  \quad ,
\eeqy
\beqy\label{33d}
G_\ell^\prime > -(2\ell+1)  \quad .
\eeqy
\emlet
The Landau parameters $F_0$ and $F_0^\prime$ are related to the usual compression modulus 
\beqy
\label{34}
K_v=\frac{3\hbar^2 k_{{\rm F}0}^2}{M_s^*}(1+F_0)  \quad ,
\eeqy
and symmetry energy
\beqy
\label{35}
J=\frac{\hbar^2 k_{{\rm F}0}^2}{6 M_s^*}(1+F_0^\prime)\quad ,
\eeqy
respectively. 
The conditions on $G_0$ and $G_0^\prime$ ensure that SNM is stable against small 
fluctuations of isoscalar and isovector spin densities respectively. These Landau 
parameters can be expressed in terms of the spin asymmetry coefficient, 
defined by
\beqy
\label{36}
a_\sigma \equiv \frac{1}{2}
\frac{\partial^2 e}{\partial I_\sigma^2}\biggr\vert_{I_\sigma = 0} = 
\frac{\hbar^2 k_{{\rm F}0}^2}{6 M_s^*}(1+G_0)\quad ,
\eeqy
and the spin-isospin asymmetry coefficient, defined by
\beqy
\label{37}
a_{\sigma\tau} \equiv \frac{1}{2}\frac{\partial^2 e}{\partial I_{\sigma\tau}^2}\biggr\vert_{I_{\sigma\tau} = 0}=\frac{\hbar^2 k_{{\rm F}0}^2}{6 M_s^*}(1+G_0^\prime)\quad ,
\eeqy
where $I_{\sigma\tau}\equiv s_1/\rho=(\rho_{n\uparrow}-\rho_{n\downarrow}-
\rho_{p\uparrow}+\rho_{p\downarrow})/\rho$. 
Using the Skyrme functional, the Landau parameters $G_0$ and $G_0^\prime$ 
are given by
\beqy
\label{38}
G_0=2 N_0\Biggl[C_0^s+C_0^T k_{{\rm F}0}^2\Biggr]\quad ,
\eeqy
\beqy
\label{39}
G_0^\prime=2N_0\Biggl[C_1^s+C_1^T k_{{\rm F}0}^2\Biggr]\quad .
\eeqy
The stability of SNM at any density thus requires
\beqy
\label{40}
C_t^T\geq 0 \quad .
\eeqy
These two conditions entail Eq.~(\ref{28}). Note that Landau's stability conditions 
allow one of the coefficients $C_t^s$ to be negative provided their sum remains 
positive.

Landau's stability conditions, Eqs.~(\ref{24b}),(\ref{33c}) and (\ref{33d}),
guarantee that the unpolarized state is \emph{locally} stable (metastable)
against \emph{small} fluctuations of the spin and spin-isospin polarizations. 
But this criterion does not necessarily imply that the unpolarized state is the ground state, i.e. 
the state with the lowest energy. In particular, the ground state could still be polarized
with finite values of $I_\sigma$ and $I_{\sigma\tau}$. Moreover we have only considered so 
far the two limiting cases of SNM and NeuM. However, the outer core of neutron stars is formed 
of isospin asymmetric nuclear matter whose composition varies with depth. We thus need a more 
general stability criterion. 
 
\subsection{General stability criterion}
\label{sec3B}

Asymmetric nuclear matter is stable with respect to \emph{any} degree of 
spin and spin-isospin polarizations whenever the energy density 
$\mathcal{E}^{\rm pol}$ of the polarized state is larger than the energy 
density $\mathcal{E}^{\rm unpol}$ of the unpolarized state (for a given density $\rho$). 
Using Eqs.~(\ref{7}), (\ref{17}) and (\ref{18}) we find 
\beqy\label{41}
\mathcal{E}^{\rm pol}=\sum_{q,\sigma} \frac{\hbar^2}{2M_{q\sigma}^*}\tau_{q\sigma}
+C_0^s s^2 +C_1^s (s_n-s_p)^2+C_0^\rho \rho^2 +C_1^\rho (\rho_n-\rho_p)^2
\eeqy
which for unpolarized matter (i.e. $s_q=0$, $T_q=0$) yields
\beqy\label{42}
\mathcal{E}^{\rm unpol}=\sum_{q} \frac{\hbar^2}{2M_{q}^*}\tau_{q}
+C_0^\rho \rho^2 +C_1^\rho (\rho_n-\rho_p)^2\quad ,
\eeqy
with 
\beqy\label{43}
\tau_q=\frac{3}{5}(3\pi^2)^{2/3}\rho_q^{5/3}\quad .
\eeqy
The difference can thus be expressed as
\beqy
\label{44}
\mathcal{E}^{\rm pol}-\mathcal{E}^{\rm unpol} &=& \sum_q \frac{\hbar^2}{2M_q^*}(\tau^{\rm pol}_q-\tau_q)
+C_0^s s^2 +C_1^s (s_n-s_p)^2 \nonumber \\
& & +C_0^T sT + C_1^T (s_n-s_p)(T_n-T_p)
\eeqy
where $\tau_q^{\rm pol}=\tau_{q\uparrow}+\tau_{q\downarrow}$ is the nucleon kinetic density 
in the polarized phase. The absolute stability of the unpolarized phase can be insured by 
requiring each term be separately positive so that $\mathcal{E}^{\rm pol}>\mathcal{E}^{\rm unpol}$. 
Now the first term in Eq.~(\ref{44}) is always positive since mechanical 
stability requires $M_q^*\geq0$ and the Pauli exclusion principle implies that $\tau_q^{\rm pol}>\tau_q$. 
Let us also remark that $(s_n-s_p)(T_n-T_p)\geq0$ because $\tau_{q\sigma}$ increases monotonically with 
$\rho_{q\sigma}$. The following constraints
\bmlet 
\beqy
\label{45a}
C_t^s\geq0\quad ,
\eeqy
and 
\beqy
\label{45b}
C_t^T\geq0\quad ,
\eeqy
\emlet 
therefore guarantee the absence of any spin-ordered phase transitions in asymmetric nuclear matter.  
It is readily seen from Eqs.~(\ref{27}), (\ref{38}) and (\ref{39}) that these inequalities 
enforce Landau stability conditions, Eq.~(\ref{24b}) in NeuM and Eqs.~(\ref{33c}) and (\ref{33d}) in 
SNM. Since Eqs.~(\ref{45a}) and (\ref{45b}) ensure the stability of asymmetric nuclear matter, they  
obviously prevent a ferromagnetic collapse of NeuM as can be seen from Eq.~(\ref{15}) remembering that 
$C_0^\tau+C_1^\tau\geq 0$ as a consequence of $M_n^*\geq0$. 

\subsection{Anomalous behavior of the entropy} 
\label{sec3D}

We have seen that the stability of nuclear matter requires that $C_t^T\geq0$. 
However, these coefficients cannot take arbitrary values. From Eq.~(\ref{4}), large positive 
values of $C_t^T$ translate into large negative values of $C_t^J$ which, in certain circumstances, 
can lead to instabilities in finite nuclei whose consequence is a major rearrangement of the 
single-particle spectrum~\cite{les07}. We will now show that these coupling constants can 
be further constrained by requiring the stability of nuclear matter with respect to any degree of 
spin and spin-isospin polarizations at non-zero temperatures. 

It was shown in Refs.~\cite{rios05, isa10} that not only 
do Skyrme forces predict a ferromagnetic transition in NeuM above a certain critical density, 
but worse this density decreases with increasing temperature due to an anomalous behavior of the 
entropy. This argument can be easily transposed to asymmetric nuclear matter as follows. 
At low temperatures (compared to nucleon Fermi energies), the difference between the entropy 
density $\mathcal{S}^{\rm pol}$ of the polarized state and the entropy density $\mathcal{S}^{\rm unpol}$ 
of the unpolarized state is approximately given by 
\beqy\label{46}
\mathcal{S}^{\rm pol}-\mathcal{S}^{\rm unpol}=\sum_{q,\sigma}\frac{\pi^2 T M_q^*\rho_q}{2\hbar^2 k_{{\rm F}q}^2}\left[\frac{M_{q\sigma}^*}{M_q^*}\left(\frac{2\rho_{q\sigma}}{\rho_q}\right)^{1/3}-1\right]\quad .
\eeqy
Now because the polarized phase is more ordered than the unpolarized phase, 
its entropy according to Boltzmann's definition should thus be lower, i.e. 
$\mathcal{S}^{\rm pol}<\mathcal{S}^{\rm unpol}$ as found in realistic calculations~\cite{bomb06,bord08b,big09,mod10}.  
Since this should be true for any isospin asymmetry, we find from Eq.~(\ref{46})
\beqy\label{47}
\sum_\sigma \frac{M_{q\sigma}^*}{M_q^*}\left(\frac{\rho_{q\sigma}}{\rho_q}\right)^{1/3}< 2^{2/3}\, .
\eeqy
This condition reduces to that of Ref.~\cite{rios05} in the limiting case of fully polarized NeuM. 
Equation~(\ref{47}) can be equivalently expressed as ($q^\prime\neq q$)
\beqy\label{47a}
\frac{(1+I_{\sigma q})^{1/3}}{1+\Xi I_{\sigma q}-\Upsilon I_{\sigma q^\prime}}+\frac{(1-I_{\sigma q})^{1/3}}{1-\Xi I_{\sigma q}+\Upsilon I_{\sigma q^\prime}}< 2\quad ,
\eeqy
with $I_{\sigma q}=(\rho_{q\uparrow}-\rho_{q\downarrow})/\rho_q$,  
\beqy\label{47b}
\Xi=(C_0^T+C_1^T)\rho_q \frac{2 M_q^*}{\hbar^2}\quad ,
\eeqy
\beqy\label{47c}
\Upsilon=(C_0^T-C_1^T)\rho_{q^\prime} \frac{2 M_{q}^*}{\hbar^2}\quad .
\eeqy
We have found numerically that the inequalities~(\ref{47a}) can be satisfied for any degree of spin and 
spin-isospin polarizations, i.e. $0<|I_{\sigma q}|,|I_{\sigma q^\prime}|\leq 1$, provided
\bmlet
\beqy\label{47d}
\Xi_1 \leq \Xi \leq \Xi_2 \quad ,
\eeqy
\beqy\label{47e}
\Upsilon=0 \, \quad ,
\eeqy
\emlet
with $\Xi_1\simeq -0.21$ and $\Xi_2\simeq 0.54$. We have also found solutions of (\ref{47a}) for 
$|\Upsilon|>\Upsilon_c(\Xi)>0$. But it can be seen from Eq.~(\ref{47c}) that such solutions cannot 
exist for all densities and must therefore be excluded. Inserting Eq.~(\ref{47b}) in Eq.~(\ref{47d}) 
using Eq.~(\ref{18}) yields
\bmlet
\beqy\label{47f}
\rho_q \left[(C_0^T+C_1^T)-\Xi_2 (C_0^\tau+C_1^\tau)\right]- \Xi_2 \rho_{q^\prime} (C_0^\tau-C_1^\tau) \leq \Xi_2 \frac{\hbar^2}{2M_q}\quad ,
\eeqy
\beqy\label{47g}
\rho_q \left[(C_0^T+C_1^T)-\Xi_1 (C_0^\tau+C_1^\tau)\right]- \Xi_1 \rho_{q^\prime} (C_0^\tau-C_1^\tau) \geq \Xi_1 \frac{\hbar^2}{2M_q}\quad .
\eeqy
\emlet
The terms in $\rho_{q^\prime}$ always satisfy the above inequalities. This is a consequence 
of the positivity of $M_q^*$ for any density and isospin asymmetry which requires that 
$C_0^\tau+C_1^\tau\geq 0$ and $C_0^\tau-C_1^\tau\geq 0$, as can be seen from Eq.~(\ref{18}). 
The conditions (\ref{47f}) and (\ref{47g}) can be ensured for any density $\rho_q$ by imposing 
that the associated terms be respectively negative and positive leading to
\beqy\label{48b}
\Xi_1 (C_0^\tau+C_1^\tau)\leq C_0^T+C_1^T\leq \Xi_2 (C_0^\tau+C_1^\tau)\, \quad .
\eeqy
On the other hand, Eq.~(\ref{47e}) implies 
\beqy\label{48a}
C_0^T=C_1^T \, \quad .
\eeqy
Combining these inequalities with Eqs.~(\ref{45b}), we arrive at the following restrictions
\beqy\label{49}
C_0^T=C_1^T\, , \quad 0 \leq C_t^T \leq \frac{1}{2}\Xi_2 (C_0^\tau+C_1^\tau)\, \quad .
\eeqy
Eqs.~(\ref{49}) guarantee that asymmetric nuclear matter remains unpolarized at finite temperature $T$
since the free energy density of the polarized phase, defined 
by $\mathcal{F}^{\rm pol}=\mathcal{E}^{\rm pol}-T\mathcal{S}^{\rm pol}$, is always higher than the 
free energy density $\mathcal{F}^{\rm unpol}=\mathcal{E}^{\rm unpol}-T\mathcal{S}^{\rm unpol}$ 
of the unpolarized phase.

\section{Stability of Skyrme forces revisited}
\label{sec4}

\begin{figure}
\centerline{\epsfig{figure=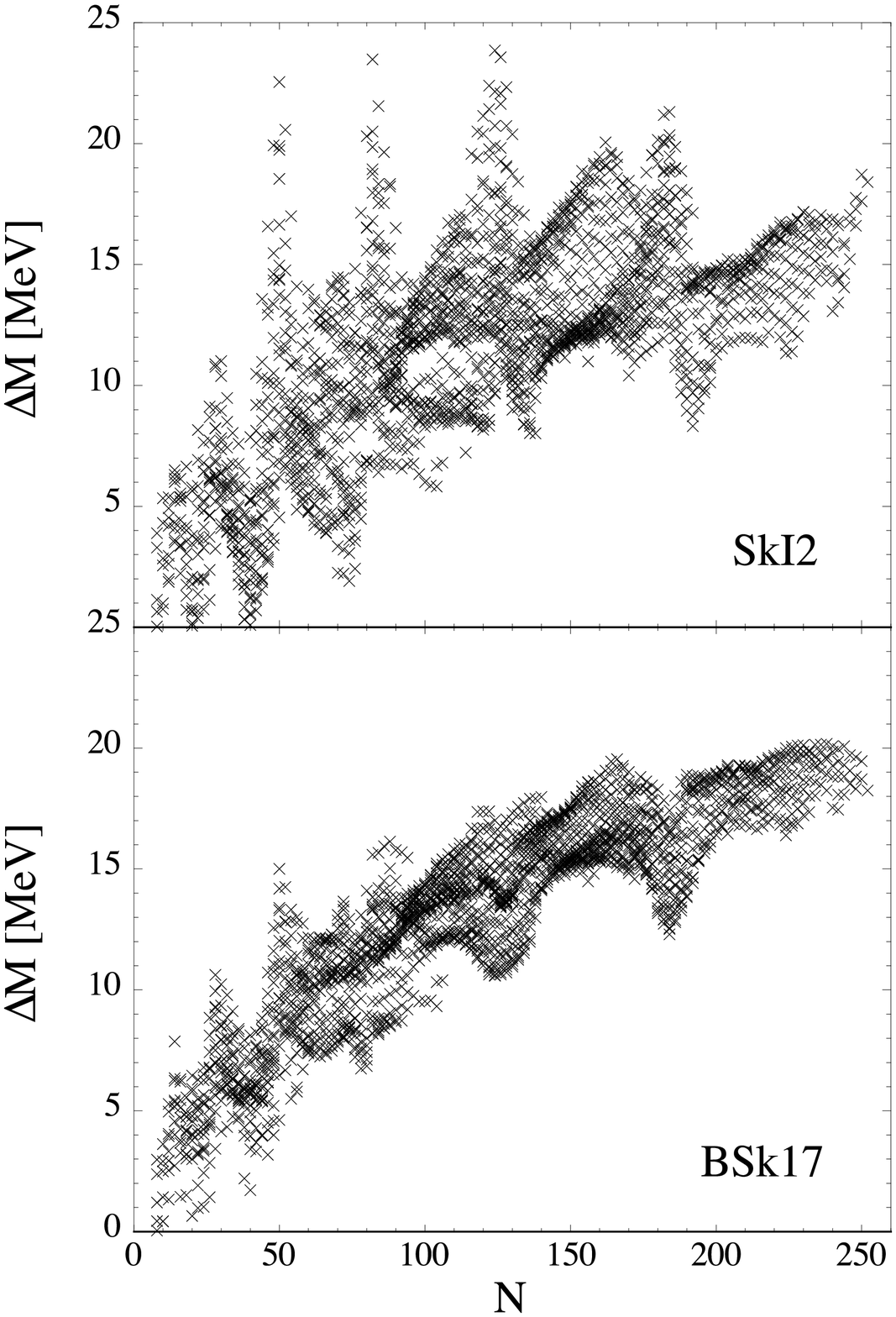,height=10.0cm,angle=0}}
\caption{Differences between the HFB energies estimated with and without 
the $J^2$-terms for two Skyrme forces SkI2 (upper panel) and BSk17 (lower panel) 
for all even-even nuclei with $Z,N>8$ and $Z<110$ lying between the proton and 
neutron driplines.}
\label{fig1}
\end{figure}

Conventional Skyrme forces have been shown to predict various spin and spin-isospin instabilities 
in nuclear matter~\cite{vida84, kw94, mar02, rios05, isa06, cgp09, isa10}. We have seen in the previous section 
that for a nuclear functional given by Eqs.~(\ref{2a}),(\ref{2b}) and (\ref{2c}) the stability of asymmetric 
nuclear matter at any temperature can be ensured by imposing Eqs.~(\ref{45a}) and (\ref{49})
[the constraint proposed in Ref.~\cite{kw94}, Eq.~(\ref{6}) and more generally Eq.~(\ref{15}), prevents a collapse 
of polarized NeuM, but does not forbid a ferromagnetic transition]. 
While the coefficients $C_t^s$ are generally positive (at least for not too high densities), 
standard Skyrme forces yield negative values of at least one of the couplings constants $C_t^T$. The origin of 
the instabilities can therefore be traced back to the time-odd terms $\pmb{s_t}\cdot\pmb{T_t}$, which are related 
to the time-even terms $\sum_{\mu,\nu}J_{t,\mu\nu}J_{t,\mu\nu}$ due to gauge invariance~(\ref{4}). Since the seminal 
work of Vautherin and Brink~\cite{vb72}, it is commonly taken for granted that the spin-current tensor (which is usually 
approximated by the spin-current vector $\pmb{J_q}$) is small in nuclei, and most Skyrme parametrizations therefore neglect them. 
We have tested this assumption by computing the HFB energies with and without the $J^2$ and $J_q^2$ terms (denoted respectively by 
$E_{\rm HFB}$ and $E^0_{\rm HFB}$) for all even-even nuclei with $Z,N>8$ and $Z<110$ lying between the proton and neutron 
drip lines (Note that when the $J^2$ and $J_q^2$ terms are 
included, the associated time-odd terms in $C_t^T$ play a role in the exact treatment of the masses of odd nuclei, but not 
in the equal-filling approximation~\cite{pmr08}, which we adopt here, as in all our previous papers). 
The differences $\Delta M\equiv E_{\rm HFB}-E^0_{\rm HFB}$ are shown in Fig.~\ref{fig1} for the Skyrme parametrization 
BSk17~\cite{gcp09, gcp09b} which was originally fitted with the $J^2$ and $J_q^2$ terms, and for SkI2~\cite{skI} which was not. 
The impact of the $J^2$ and $J_q^2$ terms is quite large, reaching about 20 MeV for the heaviest nuclei. The impact of dropping or 
including the $J^2$ and $J_q^2$ terms is logically found to be correlated to the amplitude of the $C_t^T=-C_t^J$ coupling constants, 
especially $C_0^T$. For instance, in the case of the SLy4~\cite{sly4} interaction ($C_0^T=-17.21$~MeV~fm$^5$), the HFB energy is 
affected by no more than 5 MeV, while for SkO~\cite{skO} ($C_0^T=-220.54$~MeV~fm$^5$) values up to 30~MeV can be reached. Adding or 
removing the $J^2$ and $J_q^2$ terms {\it a posteriori} without refitting all the parameters of the force can thus lead to significant 
errors. However in all previous studies of spin and spin-isospin instabilities in nuclear matter~\cite{vida84, kw94, mar02, agra04, 
rios05, isa06,gar08,mar09a,cgp09,isa10}, the time-odd terms $\pmb{s_t}\cdot\pmb{T_t}$ were taken into account whereas the Skyrme forces 
were generally fitted without the $J^2$ and $J_q^2$ terms. This treatment not only violates gauge symmetry but also introduces 
inconsistencies in the residual interaction hence in the Landau parameters (see the discussion in Section III of Ref.~\cite{bend02} and 
also in Section 5D of Ref.~\cite{les07}). 

We have therefore reexamined the stability of several Skyrme parametrizations for which the 
$J^2$ and $J_q^2$ terms were not included in the fit: SGII~\cite{sg2}, SLy4~\cite{sly4}, 
SkI1-SkI5~\cite{skI}, SkO~\cite{skO} and LNS~\cite{lns}.  
The parametrization SGII~\cite{sg2} was constructed in order to improve the Landau 
parameters $G_0$ and $G_0^\prime$ and the description of Gamow-Teller 
resonances in nuclei. 
The Skyrme Saclay-Lyon forces and especially the parametrization SLy4~\cite{sly4}, 
have been widely used not only in nuclear physics, but also in neutron-star studies 
because these forces were constrained to reproduce a realistic neutron-matter equation 
of state. 
The SkI~\cite{skI} forces were all constrained (except for SkI1) to reproduce the isotopic 
shifts of the root mean square charge radii of neutron rich Pb and Ca nuclei. 
Forces SkI3 and SkI4 were constructed with non-standard spin-orbit couplings. For the 
parametrization SkI5, the $^{16}$O ground-state data were excluded from the fit. 
We have also included the parametrization SkO~\cite{skO} from the same group. 
The parametrization LNS~\cite{lns} was fitted to Brueckner calculations. 
The Landau parameters in SNM and in NeuM calculated at saturation density $\rho_0$, 
with and without the terms in $C_t^T$ are shown in Table~\ref{tab1}. 
For comparison we have also indicated the predictions from Brueckner-Hartree-Fock calculations in 
SNM~\cite{zuo03} and from realistic calculations based on the renormalization group 
approach in NeuM~\cite{sfb03}. As can be seen in Table~\ref{tab1}, setting $C_t^T=0$ in 
Eqs.~(\ref{27}), (\ref{33c}) and (\ref{33d}) tends to reduce the discrepancies between the different 
Skyrme functionals and generally leads to a better agreement with realistic calculations, especially for 
$G_0^\prime$. In particular, the new values of $G_0^\prime$ lie closely inside the empirical range 
of $1.0\pm0.1$ deduced in Ref.~\cite{bor84} from the analysis of Gamow-Teller 
resonances and magnetic-dipole modes in finite nuclei. The improvement is quite 
spectacular for the parametrization SkI1. In the case of LNS, setting $C_0^T=0$ actually
deteriorates the value of the Landau parameter $G_0$ since the latter was directly 
fitted to the value obtained from realistic calculations. 
Table~\ref{tab2} shows the critical densities of the spin-ordered phase transitions 
according to Landau's stability criterion. It can be seen that dropping the terms $\pmb{s_t}\cdot\pmb{T_t}$ 
eliminates the instabilities in almost all Skyrme forces. This prescription is also consistent 
with Eqs.~(\ref{49}) and therefore prevents an anomalous behavior of the entropy thus ensuring the stability of 
nuclear matter for any temperatures. 

Moreover, setting $C_t^T=0$ is the only prescription which guarantees
the Landau stability conditions of Eqs.~(\ref{24b}), (\ref{33c}) and (\ref{33d}) at any density
both for $\ell=0$ and $\ell=1$. 
Indeed the Landau parameters $G_1$, $G_1^\prime$ in SNM and $G_1^{\rm NeuM}$ in NeuM, are given by
\beqy\label{50}
G_1=-2 N_0 C_0^T k_{{\rm F}0}^2 \quad ,
\eeqy
\beqy\label{51}
G_1^\prime=-2 N_0 C_1^T k_{{\rm F}0}^2 \quad ,
\eeqy
\beqy\label{52}
G_1^{\rm NeuM}=-2 N k_{\rm F}^2 (C_0^T+C_1^T) \quad .
\eeqy
Requiring $G_1\geq-3$, $G_1^\prime\geq-3$ and $G_1^{\rm NeuM}\geq-3$ for any density 
thus leads to $C_t^T\leq0$. Combining these inequalities with Eqs.~(\ref{40}) yields
$C_t^T=0$. Adopting these particular values tends to be supported by the following basic sum 
rules of Landau Fermi liquid theory~\cite{frim79}
\bmlet
\beqy\label{53a}
S_1=\sum_\ell \frac{F_\ell}{1+F_\ell/(2\ell+1)}+
\frac{F^\prime_\ell}{1+F^\prime_\ell/(2\ell+1)}  \nonumber \\
+\frac{G_\ell}{1+G_\ell/(2\ell+1)}+
\frac{G^\prime_\ell}{1+G^\prime_\ell/(2\ell+1)}  
=0   \quad ,
\eeqy
and
\beqy\label{53b}
S_2=\sum_\ell \frac{F_\ell}{1+F_\ell/(2\ell+1)}-
3\frac{F^\prime_\ell}{1+F^\prime_\ell/(2\ell+1)}   \nonumber \\
-3\frac{G_\ell}{1+G_\ell/(2\ell+1)}+
9\frac{G^\prime_\ell}{1+G^\prime_\ell/(2\ell+1)} 
=0     \quad .
\eeqy
\emlet
Even though Skyrme forces generally violate these sum rules, the prescription $C_t^T=0$
significantly improves the second sum rule as can be seen in Table~\ref{tab3}. 
It is quite remarkable that dropping the terms $\pmb{s_t}\cdot\pmb{T_t}$ not only removes 
all kinds of instabilities in nuclear matter but also improves the internal consistency of 
the nuclear functional. Nevertheless with this prescription, the Landau parameters 
$G_1$, $G_1^\prime$ and $G_1^{\rm NeuM}$ all vanish leading to unrealistic effective masses in polarized 
matter. Indeed, according to Eqs.~(\ref{17}) $M^*_{q\uparrow}=M^*_{q\downarrow}=M_q^*$ which 
obviously holds in the limit of vanishing spin polarizations but is otherwise contradicted 
by realistic calculations~\cite{bord08b,bomb06,sam07,big09,mod10}. In particular, these calculations 
indicate that in polarized NeuM $M_{n\uparrow}>M_{n\downarrow}$ whenever $\rho_\uparrow>\rho_\downarrow$. 
Imposing the less stringent stability conditions~(\ref{49}) leads to a splitting of effective masses but 
with a wrong sign. This deficiency calls for further extensions of existing Skyrme functionals.

\begin{figure}
\centerline{\epsfig{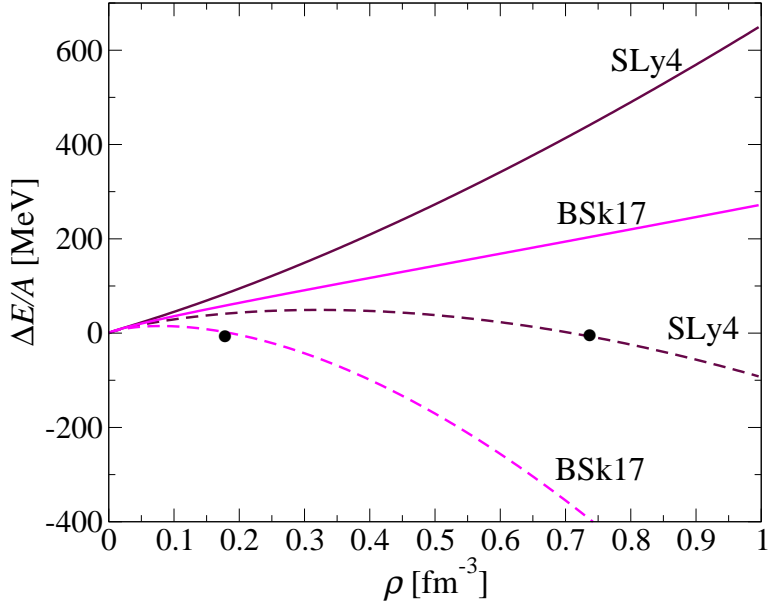}}
\caption{Difference between the energy per particle in fully polarized neutron matter 
and in unpolarized neutron matter for two Skyrme forces SLy4 and BSk17, with (dashed line)
and without (solid line) the $J^2$ and $J_q^2$ terms and their time-odd part. The black dots
indicate the densities at which the difference vanishes.}
\label{fig2}
\end{figure}

In the discussion above, we have implicitly adopted the point of view of the nuclear EDF
 theory~\cite{drut10} that the different terms appearing in Eqs~(\ref{2b}) and (\ref{2c}) 
can be \textit{a priori} considered as independent from each other (apart from the requirements of gauge invariance). 
It is therefore perfectly legitimate to set $C_t^J=-C_t^T\equiv 0$. However in the framework of effective 
forces, the coupling constants are uniquely determined by the parameters of the force. 
In particular, the coefficients $C_t^s$ and $C_t^T$ are now given by
\bmlet
\beqy
\label{54a}
C_0^s=-\frac{1}{4}t_0\left(\frac{1}{2}-x_0\right)-\frac{1}{24}t_3\left(\frac{1}{2}-x_3\right)\rho^\alpha
\eeqy
\beqy
\label{54b}
C_1^s=-\frac{1}{8}t_0-\frac{1}{48}t_3\rho^\alpha
\eeqy
\beqy
\label{54c}
C_0^T=-\frac{1}{8}\biggl[t_1\left(\frac{1}{2}-x_1\right)-t_2\left(\frac{1}{2}+x_2\right)\biggr]
\eeqy
\beqy
\label{54d}
C_1^T=-\frac{1}{16}(t_1-t_2)\, . 
\eeqy
\emlet
We have therefore studied the stability of the few Skyrme parametrizations which were fitted with 
the $J^2$ and $J_q^2$ terms: SkP~\cite{doba84}, SLy5~\cite{sly4}, SkO$^\prime$~\cite{skO}, SkX~\cite{bro98} and BSk17~\cite{gcp09, gcp09b}.
The parametrization SkP, which was specifically designed to be used both in the particle-hole channel and in the 
particle-particle channel, is still used nowadays. The forces SLy5 and SkO$^\prime$ were 
fitted following the same protocol as SLy4 and SkO respectively, but they include the contribution of the 
$J^2$ and $J_q^2$ terms. The force SkX~\cite{bro98} was constructed in an attempt to improve the 
description of single-particle energies. BSk17 is the force underlying our nuclear mass model HFB-17, 
based on the Hartree-Fock-Bogoliubov method~\cite{gcp09, gcp09b}. With this model we were able 
to fit with an rms deviation of 0.581 MeV the 2149 measured masses of nuclei with $N$ and $Z \ge 8$ 
given in the 2003 Atomic Mass Evaluation~\cite{audi03}, while at the same time constraining the underlying 
Skyrme force to fit properties of SNM and NeuM, as determined by many-body 
calculations using realistic potentials. The values of the Landau parameters in SNM and in NeuM are shown in 
Table~\ref{tab4} and the critical densities for the onset of instabilities are shown in Table~\ref{tab5}. 
For those few Skyrme forces which include the $J^2$ and $J_q^2$ terms, nuclear matter is therefore unstable 
because of the tight correlations between the different coupling constants in the energy density. 
In order to illustrate the impact of the $J^2$ and $J_q^2$ terms and their time-odd counterparts on the 
stability of nuclear matter, we have plotted in Fig.~\ref{fig2} the difference between the energy per particle
in fully polarized NeuM and in unpolarized NeuM for the parametrizations SLy4 and BSk17. Both have been fitted 
to a realistic equation of state of NeuM, but BSk17 includes the  $J^2$ and $J_q^2$ terms while SLy4 does not.
Removing all instabilities requires that we impose $C_t^s\geq0$ and $C_t^T=0$. Since the first term in $t_0$ of the Skyrme 
force is generally associated with the long-range attractive part of the nucleon-nucleon interaction 
while the density-dependent term in $t_3$ is related to the strongly repulsive short-range
part, the coupling constant $C_0^s$ can be made positive for any density by choosing 
$x_0<1/2$ and $x_3>0$. With $t_0<0$ and $t_3>0$, the coefficient $C_1^s$ will be positive, 
at least for not too high densities. Spin- and spin-isospin instabilities thus generally 
arise mainly from the coupling constants $C_0^T$ and $C_1^T$, which in turn are generated 
by the momentum-dependent terms in $t_1$ and $t_2$. 
Using Eqs.~(\ref{54c}) and (\ref{54d}), the conditions $C_t^T=0$ entail $t_1=t_2$ and 
$x_1=-x_2$. Imposing these constraints would leave no degree of freedom 
for adjusting surface properties of nuclei, which also depend on the momentum 
dependent $t_1$ and $t_2$ terms through the coupling constants $C_t^{\Delta\rho}$. 
This would also have an impact on the coupling constants $C_t^\tau$ which 
determine the nucleon effective masses, Eq.~(\ref{18}). There is little doubt 
that such a force would yield poor results when applied to nuclei. 
Since thermal effects on the spin polarization are rather small for temperatures found in 
protoneutron stars and supernova cores~\cite{rios05}, one may be tempted to 
require the stability of cold nuclear matter only. But even in this case, it was shown 
in Refs.~\cite{mar02,agra04} that it is not possible to avoid spurious transitions to spin-ordered 
phases in nuclear matter above 2--3 times saturation density, and at the same time giving 
reasonable properties of SNM. We have found that the critical densities above which instabilities
occur are even lower when more nuclear data are included in the fit of the effective interaction.
In particular, conventional Skyrme forces fitted to essentially all experimental nuclear mass data 
predict a ferromagnetic transition in NeuM at a density slightly above saturation 
density~\cite{cgp09} (see also Table~\ref{tab5}). 

The stability of cold nuclear matter can only be restored by including additional components 
in the Skyrme interaction, thereby inducing new terms in the energy density. Two different extensions 
have been recently proposed. Margueron and Sagawa~\cite{mar09a} considered extended Skyrme forces 
with two new $t_3$ like terms depending on the nucleon spin densities $\pmb{s_q}$ of the form
\beqy\label{55}
\frac{1}{6}t_3^s(1+x_3^s P_\sigma) s(\pmb{r})^2\,\delta(\pmb{r}_{ij})
+\frac{1}{6}t_3^{st}(1+x_3^{st} P_\sigma) s_1(\pmb{r})^2\,\delta(\pmb{r}_{ij})\quad .
\eeqy
In the energy density, Eqs.~(\ref{2b}) and (\ref{2c}), these new terms modify the coefficients 
$C_t^s$. The additional parameters were adjusted so as to ensure the 
Landau stability conditions $G_0>-1$, $G_0^\prime>-1$ and $G_0^{\rm NeuM}>-1$. 
The nuclear mass model HFB-17~\cite{gcp09, gcp09b} was thus refitted with these new terms~\cite{marg09}. 
With this extended Skyrme force called BSk17st, it was possible to maintain the quality
of the HFB-17 mass model, and at the same time the Landau parameters were adjusted 
so as to remove the spin and spin-isospin instabilities present in the original force 
BSk17. Unfortunately instabilities were still found for finite spin and 
spin-isospin polarizations~\cite{marg09}. The reason is that terms of the form given by 
Eq.~(\ref{55}), do not change the coefficients $C_t^T$ and consequently, Eq.~(\ref{44})
is not guaranteed to remain positive for any spin and spin-isospin polarizations. 
Moreover, as noted in Ref.~\cite{marg09} the contributions of Eq.~(\ref{55}) to the energy 
density cancel in fully polarized NeuM so that BSk17st still predicts a ferromagnetic collapse of 
NeuM as BSk17 does. The extension of Ref.~\cite{mar09a} does not affect the coefficients
$C_t^T$ hence also the effective masses of spin-up and spin-down nucleons are not affected, as can be seen from
Eq.~(\ref{17}). This 
means that if the original Skyrme force violates the constraint~(\ref{47}), this will 
still be the case for the extended version of this force.

Alternatively, instabilities can be avoided by introducing into the force, density-dependent
generalizations of the usual $t_1$ and $t_2$ terms of the form~\cite{cgp09}
\beqy\label{56}
\frac{1}{2} t_4(1+x_4 P_\sigma)\frac{1}{\hbar^2}\left[p_{ij}^2\,\rho(\pmb{r})^\beta
\delta({\pmb{r}_{ij}}) +\delta({\pmb{r}_{ij}})\rho(\pmb{r})^\beta\, p_{ij}^2 \right]
+t_5(1+x_5 P_\sigma)\frac{1}{\hbar^2}\pmb{p}_{ij}\cdot\rho(\pmb{r})^\gamma\delta(\pmb{r}_{ij})\,
 \pmb{p}_{ij} \, .
\eeqy
These new terms modify the coefficients $C_t^\tau$, $C_t^T$, $C_t^{\Delta\rho}$ and 
$C_t^{\Delta s}$ thus providing more flexibility to remove instabilities without 
deteriorating the fit to nuclear data. We have constructed a new nuclear mass model, labeled 
HFB-18, with such a generalized Skyrme force~\cite{cgp09}. The parameters $t_5$, $x_5$ and 
$\gamma$ were chosen in order to avoid a ferromagnetic collapse of neutron-star 
matter. For simplicity, the remaining parameters in Eq.~(\ref{56}) were fixed by the equations 
\bmlet 
\beqy\label{57a}
\beta = \gamma  \quad ,
\eeqy
\beqy\label{57b}
t_4 = -\frac{1}{3}t_5(5 + 4x_5)  \quad ,
\eeqy
\beqy\label{57c}
x_4 = -\frac{4 + 5x_5}{5 + 4x_5} \quad ,
\eeqy
\emlet
which ensure that the contributions of the new terms to the coefficients $C_t^\tau$ vanish identically.
As a result, the $t_4$ and $t_5$ terms cancel exactly in unpolarized homogeneous nuclear matter. 
This new model yields almost as good a mass fit as our previous model HFB-17, 
with the advantage that NeuM matter is now stable with respect to any degree of spin 
polarizations. Even thouth this new force still predicts an isospin instability in SNM, this does not affect 
the interior of neutron stars which is now unpolarized. Moreover we have found 
that this isospin instability can be easily removed if the conditions~(\ref{57a})--(\ref{57c}) are released, 
without deteriorating the quality of the mass fit~\cite{gcp10}. However 
we did not succeed in constructing a nuclear mass model that satisfies Eq.~(\ref{47}).
As a consequence, nuclear matter could still become unstable at finite temperatures even though
no phase transitions occur at zero temperature, as shown in Ref.~\cite{rios05}. 

One might be tempted to enforce the stability conditions $C_t^T=0$ by adding a zero-range tensor force to the conventional Skyrme 
interaction~(\ref{5}) with suitable adjustments of the parameters, like the parametrization T22 of 
Ref.~\cite{les07}. Unfortunately, a tensor force introduces new terms in the functional which also 
affect the stability of nuclear matter~\cite{back79}. The stability of 41 different Skyrme interactions having a tensor 
component has been recently studied in Ref.~\cite{cao10}. In particular, the recent Skyrme forces from the 
Saclay-Lyon group~\cite{les07} which include tensor forces and which were fitted following the same protocol 
as the older SLy family~\cite{sly4}, still predict various spin and spin-isospin instabilities. This is 
notably the case for the force T22 for which $C_t^T=0$. 

\section{Conclusion}

Nuclear energy density functional theory has been traditionally restricted to very
specific phenomenological semi-local functionals of the form given by Eqs.~(\ref{2a})--(\ref{2c}), based on effective 
forces~\cite{bhr03,sto07}. However the use of effective forces introduces tight correlations between different terms of the 
functional, which can generate various kinds of instabilities. 
In particular, the time-odd terms $\pmb{s_t}\cdot\pmb{T_t}$ induced by the momentum-dependent part of Skyrme forces 
(which contribute also to the coupling constants $C_t^\tau$, $C_t^{\Delta\rho}$ and $C_t^{\Delta s}$) are responsible 
for spurious spin and spin-isospin instabilities in infinite homogeneous nuclear matter at densities 
encountered in the interior of neutron stars. In some cases, instabilities arise in symmetric nuclear
matter below saturation densities and could thus also contaminate calculations in finite nuclei
(Note that the coupling constants $C_t^{\Delta\rho}$ alone were found to drive finite-size instabilities~\cite{kor10}). 
These correlations between different parts of the nuclear energy density functional hamper the 
development of more accurate functionals since adding one term in the effective force can induce
several new terms in the functional. Moreover, the coupling constants of the time-odd terms are generally 
not directly fitted to experimental data but are calculated {\it a posteriori} using the parameters of Skyrme force. 
However, there is no guarantee that the effects associated with the time-odd terms will be correctly described
in this way. As shown in Refs.~\cite{mar02,agra04} it is not possible to avoid spurious transitions to spin-ordered 
phases in nuclear matter above 2--3 times saturation density. The critical densities above which these instabilities
occur, decrease when more nuclear data are included in the fit of the parameters of the Skyrme force~\cite{cgp09}. 
For instance, for our nuclear mass model HFB-17~\cite{gcp09,gcp09b}, the ground state of neutron matter becomes 
ferromagnetic above 0.17 fm$^{-3}$. 
These instabilities can be (at least partially) removed by suitable extensions of the Skyrme force, 
as proposed for instance in Refs.~\cite{mar09a, cgp09}. However an unphysical spin-ordering could still occur at finite 
temperatures thus spoiling the application of Skyrme forces to the hot nuclear matter found in protoneutron stars and 
supernova cores. Alternatively the terms $\pmb{s_t}\cdot\pmb{T_t}$ that are responsible for spin and spin-isospin instabilities 
could be canceled by suitable adjustments of an additional tensor component to the Skyrme force~\cite{les07}. 
Unfortunately a tensor force would also generate new terms in the energy density which still lead to instabilities~\cite{cao10}.

On the other hand, the concept of effective forces leads to formal inconsistencies as recently discussed in Ref.~\cite{erl10}.
Lots of efforts are now devoted to the construction of non-empirical functionals from realistic interactions directly without 
resorting to effective forces~\cite{drut10}. If one adopts the point of view that the nuclear functional 
is more fundamental than effective forces, the different terms appearing in Eqs.(\ref{2b}) 
and (\ref{2c}) can be treated independently (apart from the requirements of gauge invariance
and cancellation of self-interactions). It is therefore perfectly legitimate to 
set $C_t^J=-C_t^T\equiv 0$. Actually the $J^2$ and $J_q^2$ terms are dropped in most Skyrme forces, not only because 
of simplicity but also because it seems to be favored by global fits to nuclear data and basic nuclear matter properties~\cite{klu09}. 
Moreover, the $J^2$ and $J_q^2$ terms might even lead to instabilities in the single-particle spectra of finite nuclei, as discussed 
for instance in Ref.~\cite{les07}. 
However in all previous studies of spin and spin-isospin instabilities in nuclear 
matter~\cite{vida84, kw94, mar02, agra04, rios05, isa06,gar08,mar09a,cgp09,isa10}, the associated time-odd terms 
$\pmb{s}\cdot\pmb{T}$ and $(\pmb{s_n}-\pmb{s_p})\cdot(\pmb{T_n}-\pmb{T_p})$ have been included in the residual 
interaction thus violating gauge symmetry. We have therefore reexamined the stability of nuclear matter by setting 
$C_t^T\equiv 0$ for those Skyrme parametrizations which were fitted \emph{without} the $J^2$ and $J_q^2$ terms. 
We have found that this simple prescription not only improves the values of the Landau parameters $G_0$, $G_0^\prime$
and $G_0^{\rm NeuM}$. But this also generally removes all kinds of instabilities in asymmetric nuclear matter 
both at zero and finite temperatures. Nevertheless this prescription yields unrealistic values of the Landau parameters 
$G_1$, $G_1^\prime$ and $G_1^{\rm NeuM}$, hence also of the effective masses $M_{q\sigma}^*$ in polarized matter. 
Further improvements thus require extensions of existing Skyrme functionals.

{\it Acknowledgments}. 
This work, which was initiated by discussions with J.M. Pearson, was financially supported by 
FNRS (Belgium), Communaut\'e fran\c{c}aise de Belgique (Actions de Recherche Concert\'ees) 
and by CompStar (a Research Networking Programme of the European Science Foundation).

\begin{table}
\centering
\caption{Landau parameters $G_0$ and $G_0^\prime$ in symmetric nuclear matter and $G_0^{\rm NeuM}$ 
in neutron matter (at saturation density) for selected Skyrme forces which were fitted 
without the $J^2$ and $J_q^2$ terms. Values in parenthesis were obtained 
by setting $C_t^T=0$. The last line shows the Landau parameters 
predicted by microscopic calculations using realistic interactions: 
Ref.~\cite{zuo03} for symmetric nuclear matter and Ref.~\cite{sfb03} for neutron matter.}
\label{tab1}
\vspace{.5cm}
\begin{tabular}{|c|c|c|c|}
\hline
& $G_0$ & $G_0^\prime$ & $G_0^{\rm NeuM}$ \\ 
\hline
SGII & 0.01 (0.62) & 0.51 (0.93) & -0.07 (1.19) \\ 
SLy4 &  1.11 (1.39) & -0.13 (0.90) & 0.11 (1.27) \\
SkI1 & -8.74 (1.09) & 3.17 (0.90) & -5.57 (1.10) \\ 
SkI2 & -1.18 (1.35) & 0.77 (0.90) & -1.08 (1.24) \\ 
SkI3 & 0.57 (1.90) & 0.20 (0.85) & -0.19 (1.35) \\
SkI4 & -2.81 (1.77) & 1.38 (0.88) & -2.03 (1.40) \\
SkI5 & 0.28 (1.79) & 0.30 (0.85) & -0.31 (1.30) \\
SkO &  -4.08 (0.48) & 1.61 (0.98) & -3.17 (0.97) \\
LNS &  0.83 (0.32) &  0.14 (0.92) & 0.59 (0.91) \\ 
\hline
Realistic & 0.83 & 1.22 & 0.77 \\
\hline
\end{tabular}
\end{table}

\begin{table}
\centering
\caption{Critical densities above which nuclear matter becomes unstable
according to Landau's criterion for selected Skyrme forces which were fitted 
without the $J^2$ and $J_q^2$ terms. 
The first two column are for symmetric nuclear matter, while the last 
column is for pure neutron matter. The densities indicated in parenthesis were
obtained by setting $C_t^T=0$.}
\label{tab2}
\vspace{.5cm}
\begin{tabular}{|c|c|c|c|}
\hline
& $\rho_c(G_0)$ {\rm [fm$^{-3}$]}& $\rho_c(G_0^\prime)$ {\rm [fm$^{-3}$]}& $\rho_c(G_0^{\rm NeuM})$ {\rm [fm$^{-3}$]}\\ 
\hline
SGII & 0.44 ($\infty$) & 0.80 ($\infty$) & 0.26 (2.07) \\ 
SLy4 &  $\infty$ ($\infty$) & 0.33 ($\infty$) & 0.59 ($\infty$) \\
SkI1 & 0.04 (0.71) & $\infty$ ($\infty$) & 0.05 ($\infty$) \\ 
SkI2 & 0.14 ($\infty$) & $\infty$ ($\infty$) & 0.15 ($\infty$) \\ 
SkI3 & 0.91 ($\infty$) & 0.92 ($\infty$) & 0.37 ($\infty$) \\
SkI4 & 0.07 ($\infty$) & $\infty$ ($\infty$) & 0.09 ($\infty$) \\
SkI5 & 0.43 ($\infty$) & 1.36 ($\infty$) & 0.28 ($\infty$) \\ 
SkO &  0.07 (0.52) & $\infty$ (2.32) & 0.09 (0.67) \\
LNS & $\infty$ ($\infty$) &  0.43 ($\infty$) & 0.62 (1.38) \\ 
\hline
\end{tabular}
\end{table}

\begin{table}
\centering
\caption{Landau sum rules given by Eqs.~(\ref{53a}) and (\ref{53b}) for selected Skyrme forces which were fitted 
without the $J^2$ and $J_q^2$ terms. Values in parenthesis were obtained by setting $C_t^T=0$.}
\label{tab3}
\vspace{.5cm}
\begin{tabular}{|c|c|c|}
\hline
& $S_1$ & $S_2$ \\ 
\hline
SGII & 0.97 (0.61) & 1.13 (-0.51)\\
SLy4 & -0.31 (-0.65) & 1.52 (0.85)  \\
SkI1 & -6.71 (-0.59) & -89.2 (0.86)  \\ 
SkI2 & 6.87 (-0.71) & -20.7 (0.98) \\ 
SkI3 & -1.46 (-2.33) & 2.14 (1.84) \\
SkI4 & 1.01 (-1.23) & -11.3 (1.32) \\
SkI5 & -1.47 (-2.28) & 2.17 (1.77) \\
SkO & 3.21 (1.07) & -13.7 (0.87)  \\
LNS & 0.49 (0.63) & 3.53 (-0.04) \\ 
\hline
\end{tabular}
\end{table}

\begin{table}
\centering
\caption{Landau parameters $G_0$ and $G_0^\prime$ in symmetric nuclear matter and $G_0^{\rm NeuM}$ 
in neutron matter (at saturation density) for selected Skyrme forces which were fitted 
with the $J^2$ and $J_q^2$ terms. The last line shows the Landau parameters 
predicted by microscopic calculations using realistic interactions: 
Ref.~\cite{zuo03} for symmetric nuclear matter and Ref.~\cite{sfb03} for neutron matter.}
\label{tab4}
\vspace{.5cm}
\begin{tabular}{|c|c|c|c|}
\hline
& $G_0$ & $G_0^\prime$ & $G_0^{\rm NeuM}$ \\ 
\hline
SkO$^\prime$ & -1.62 & 0.79 & -1.43  \\ 
SLy5  & 1.09 & -0.16 & 0.09 \\ 
SkP  & -0.23 & 0.06 & -0.61 \\
SkX & -0.63 & 0.51 & -0.50 \\
BSk17 & -0.69  & 0.50  & -0.88  \\
BSk17st & -0.68 & 0.50 & 0.47 \\
BSk18 & -0.33  & 0.46  & -0.57  \\
\hline
Realistic & 0.83 & 1.22 & 0.77 \\
\hline
\end{tabular}
\end{table}

\begin{table}
\centering
\caption{Critical densities above which nuclear matter becomes unstable
according to Landau's criterion for selected Skyrme forces which were fitted 
with the $J^2$ and $J_q^2$ terms. The first two column are for symmetric nuclear matter, while the last 
column is for pure neutron matter. }
\label{tab5}
\vspace{.5cm}
\begin{tabular}{|c|c|c|c|}
\hline
& $\rho_c(G_0)$ {\rm [fm$^{-3}$]}& $\rho_c(G_0^\prime)$ {\rm [fm$^{-3}$]}& $\rho_c(G_0^{\rm NeuM})$ {\rm [fm$^{-3}$]}\\ 
\hline
SkO$^\prime$ & 0.12 & 0.97 & 0.14 \\
SLy5 &  $\infty$  & 0.33 & 0.57  \\
SkP & 0.74 & 0.30 & 0.19  \\
SkX & 0.22 & 0.40 & 0.19 \\
BSk17 & 0.21  & 0.68  & 0.17 \\ 
BSk17st & $\infty$ & $\infty$ & $\infty$ \\
BSk18 & $\infty$ & 0.62 & $\infty$ \\ 
\hline
\end{tabular}
\end{table}

\end{document}